\documentclass[prd,aps,twocolumn,reprint,floatfix,nofootinbib,superscriptaddress]{revtex4}
\pdfoutput=1
\usepackage{amssymb,graphicx}
\usepackage{amsmath}
\usepackage{multirow}
\usepackage{epsfig}
\usepackage[usenames]{color} 
\usepackage[export]{adjustbox}
\usepackage{mathtools}

\newcommand{\beqn}{\begin{eqnarray}}
\newcommand{\eeqn}{\end{eqnarray}}
\newcommand{\beq}{\begin{equation}}
\newcommand{\eeq}{\end{equation}}

\def\mphi{m_{\phi}}

\def\pt{\tilde{p}}
\def\rt{\tilde{\rho}}

\def\psib{\bar{\psi}}
\def\Phib{\overline{\Phi}}

\begin{document}

\title{Spontaneous growth of gauge fields in gravity through the Higgs mechanism}
\author{Fethi M.\ Ramazano\u{g}lu}
\affiliation{Department of Physics, Ko\c{c} University, \\
Rumelifeneri Yolu, 34450 Sariyer, Istanbul, Turkey }
\date{\today}

\begin{abstract}
We introduce gravity theories featuring spontaneously growing gauge fields where the growth is
due to the Higgs mechanism. The underlying physics is inspired
by the spontaneous scalarization phenomena in scalar-tensor theories. The tachyonic instability 
that causes the growth in spontaneous scalarization and its analog for vector fields is introduced not
as an explicit potential term, but through a scalar coupling using a Higgs field.
The resulting theories
are distinct from previous  examples of spontaneous tensorization in that they respect the gauge symmetry at 
the level of the action. Our results are valid for both Abelian and non-Abelian gauge theories, and
this is the first study of the spontaneous growth of the latter.
We discuss astrophysical implications of these theories and argue their relevance
especially in the strong gravity regime.
\end{abstract}
\maketitle

\section{Introduction}
Scalar-tensors theories are alternative theories to general relativity (GR)
where gravitation is governed by additional scalar fields
together with the metric~\cite{Fujii:2003pa}. When certain coupling
functions of these theories are appropriate, the solution with vanishing scalar
fields that corresponds to GR becomes unstable inside a neutron star (NS)~\cite{PhysRevLett.70.2220}.
Arbitrarily small scalar field perturbations exponentially grow due to a 
a tachyonic instability, but the growth eventually
stops to form a scalar cloud around the star.
The final stable field configuration can routinely attain
large amplitudes, hence deviations from GR are prominent near the NS. At the
same time the scalar dies off with distance from the star.
This presents an ideal scenario from an observational
point of view: weak-field tests of gravity are satisfied in the far field, and large deviations
in strong gravity provides a relatively easy target for gravitational
wave detection which is precision-limited~\cite{Will:2005va,0264-9381-32-24-243001}.

The central idea of spontaneous scalarization, a tachyonic instability of the
zero-field solution, has been recently generalized to other fields such
as vectors~\cite{Ramazanoglu:2017xbl}. NSs with vector clouds around them 
have similar properties to scalarized stars, hence they have the same 
appeal in terms of observations.
However the gauge symmetry we find in the
known vector fields in the universe is explicitly broken in the actions of these new theories.
This is not necessarily a problem as long as we
remain in the classical realm, but it is an unnatural quality in light of
our knowledge about fundamental fields. 

In this study we present
a new mechanism for spontaneous growth of gauge fields, Abelian or non-Abelian,
that respects the gauge symmetry at the level of the action. 
The essence of our work is a simple idea
that features the Higgs mechanism~\cite{frampton2008gauge,Srednicki:2007qs}.
The tachyon of spontaneous
scalarization arises from a local effective mass term that has the ``wrong'' sign.
This means there are modes with imaginary energy that
grow in time instead of oscillate, and form an instability.
A Higgs coupling term also generates mass for gauge fields, so 
it can also incite tachyons and lead to spontaneous growth with a similar ``wrong'' sign choice.
We name the resulting phenomena \emph{Higgs-based spontaneous growth}.
The required nonzero expectation value of the Higgs field can be obtained
through a Mexican hat potential term or the recently studied gravitational
Higgs mechanism~\cite{Coates:2016ktu}. We will present 
spontaneous growth theories for each case.

We will discuss the astrophysical relevance of Higgs-based spontaneous growth 
in comparison to what we already know about spontaneous scalarization and its
generelizations. An important point is that the equation of motion (EOM)
of the Higgs field itself is also modified in these new theories 
in a similar way to ghost-based spontaneous scalarization theories~\cite{Ramazanoglu:2017yun}. 
This has important implications for NS structure.

Sec.~\ref{sec_scalar_vector} is a summary of spontaneous scalarization and its
various generalizations.
Sec.~\ref{higgs_based} is the heart of this study where we
present Higgs-based spontaneous growth of Abelian and non-Abelian field theories.
Sec.~\ref{discussions} contains our comments on the results, comparison
to the existing literature and observational prospects of Higgs-based spontaneous
growth. We use gravitational units $G=c=1$.

\section{Spontaneous Scalarization and Its Generalizations}\label{sec_scalar_vector}
The simplest example of spontaneous growth in gravity and the inspiration for all other spontaneously
growing fields is \emph{spontaneous scalarization}~\cite{PhysRevLett.70.2220}
described by the action 
\begin{align}\label{st_action}
 \frac{1}{16\pi} &\int dV R -  \frac{1}{4\pi} \int dV \bigg[
 \overbrace{\frac{1}{2}g^{\mu \nu} \partial_{\mu} \phi  \partial_{\nu} \phi}^{T_\phi}\
 \overbrace{+ \frac{1}{2} m_{\phi}^2 \phi^2}^{V_\phi} \bigg] \nonumber \\
 &+ S_m \left[f_m, A^2(\phi) g_{\mu \nu} \right]
\end{align}
where $g_{\mu\nu}$ is the metric, $\phi$ is a real scalar field, $m_{\phi}$ is 
the parameter of the scalar potential which can be interpreted as mass in the frame of $g_{\mu\nu}$,
and $dV=d^4x \sqrt{-g}$. 
$S_m$ is the matter action, $f_m$ representing all matter degrees of freedom. 
The \emph{alternative} nature of the theory to GR arises from the nonminimal coupling in $S_m$
where matter directly interacts with the conformally scaled metric
$\tilde{g}_{\mu\nu} = A^2(\phi) g_{\mu\nu}$. The reference frame defined by this metric is 
traditionally called the Jordan frame, while that of $g_{\mu\nu}$ is the Einstein frame. We will
quickly summarize the interesting properties of this action, but a more detailed account with
our particular emphases can be found in~\cite{Ramazanoglu:2016kul}.

If the conformal scaling has the specific form
$A(\phi)=e^{\beta \phi^2/2}$ or more generally has the Taylor expansion
$A(\phi) = 1 +\beta \phi^2 + \ldots$, it turns out the $\phi=0$ solution is 
unstable in the presence of matter for low Fourier modes, i.e. any small
perturbation initially grows exponentially~\cite{PhysRevLett.70.2220}.
The reason for this can be quickly identified in the equation of motion (EOM)
\begin{align} \label{scalar_eom}
  \Box_g \phi &= \left( - 8 \pi A^4 \frac{d\left( \ln A(\phi) \right)}{d(\phi^2)} \tilde{T} + m^2_\phi \right)\phi \nonumber\\
  &\approx  ( - 4 \pi \beta \tilde{T} + m^2_\phi )\phi 
\end{align}
where $\tilde{T}$ is the trace of the stress-energy tensor in the Jordan frame,
and we linearized the equation around $\phi=0$ in the second line.
It is clear that the scalar behaves as if it has an effective mass-square
$m^2_{\rm eff}=( - 4 \pi \beta \tilde{T} + m^2_\phi )$, and $m^2_{\rm eff}<0$
for $\beta < 0$ since $\tilde{T}<0$ in most cases.
A scalar with imaginary mass is called a \emph{tachyon}.
The dispersion relation $E^2=m^2+k^2$ implies that the energies of low $k$
modes are also imaginary, hence these modes grow exponentially in time
rather than oscillate. 

When $\phi$ grows large enough,
$A^4$ factor in the EOM kills the negative mass-squared term and ensures that the growth
eventually stops. For the natural choice where $|\beta| \sim 1$, the relevant length scales mean that
NSs possess scalar clouds, but less compact stars do not. 
The scalar amplitude is generically large inside and near the star, hence
this theory promises relatively easily observable deviations from GR. On the other hand,
the scalar dies off with distance from the star in any physical solution,
ensuring agreement with weak-field tests of gravity.
These aspects of spontaneous scalarization has made it popular
for strong gravity physics and gravitational wave science.

Let us examine why we chose $A(\phi)=e^{\beta \phi^2/2}$, which will help
us understand other theories of spontaneous growth later.
When we vary the action with respect to $\phi$, the scalar
potential ($V_\phi$) contribution to the equation of motion is $m_\phi^2 \phi$, the mass-square term.
If $A(\phi)$ has a form similar to $V_\phi$, its variation also brings a
mass-square term. In short, \emph{$A(\phi)$ can be used 
to generate mass if it has the correct functional form.
This mass can be made negative, i.e. tachyonic, by choosing
the ``wrong'' sign for the coefficients in the Taylor expansion of $A$.}

Our understanding of the physics of spontaneous scalarization recently gave
rise to generalization of this phenomena in two directions. First, note that there is
nothing special about the nature of the field itself in our explanation for spontaneous 
growth. Thus, if we
have another field, e.g. a vector, and introduce a nonminimal matter coupling
that provides an effective mass term with the appropriate sign, we expect to observe
\emph{spontaneous vectorization}. This idea was realized using the
action~\cite{Ramazanoglu:2017yun}
\begin{align}\label{action_vt}
 \frac{1}{16\pi} &\int dV R -  \frac{1}{16\pi} \int dV \left[ F^{\mu\nu} F_{\mu\nu} 
 +2m_X^2 X^\mu X_\mu \right] \nonumber \\
 &+ S_m \left[f_m, A_X^2(\eta) g_{\mu \nu} \right], \ \eta =g^{\mu\nu}X_\mu X_\nu
\end{align}
where $X$ is a vector field, $F_{\mu\nu} = \nabla_\mu X_\nu -\nabla_\nu X_\mu$
and $A_X=e^{\beta_X \eta/2}$. The EOM
\begin{align} \label{eom_vt}
\nabla_\rho F^{\rho \mu} = \left(-8\pi A_X^4 \Lambda \tilde{T} +m_X^2 \right) X^\mu 
\end{align}
easily shows the imaginary effective mass, hence the tachyon and spontaneous growth.
Note the choice of $A_X$ that is inspired from spontaneous scalarization.

The second path to generalize spontaneous growth takes advantage of 
the fact that the nature of the instability is not critical either, i.e. as long as 
the zero-field solution is unstable, the field grows spontaneously. For example,
just as the tachyon effectively modifies the potential term $V_\phi$ in
Eq.~\ref{st_action}, one can have the ``wrong'' sign in the kinetic term
$T_\phi$ due to derivative coupling terms in the matter action.
Such an instability is called a \emph{ghost}, and it can be used
to construct a \emph{ghost-based spontaneous scalarization} theory.
Concretely, the action~\cite{Ramazanoglu:2017xbl}
\begin{align}\label{st_action_ghost}
 \frac{1}{16\pi} &\int dV R -  \frac{1}{16\pi} \int dV \left[ 2g^{\mu \nu} \partial_{\mu} \phi  \partial_{\nu} \phi
 + 2 m_{\phi}^2 \phi^2 \right] \nonumber \\
 &+ S_m \left[f_m, A_{\partial}^2(\kappa) g_{\mu \nu} \right] \ , \ \kappa \equiv g^{\mu\nu}\partial_\mu \phi \partial_\nu \phi
\end{align}
leads to the EOM
\begin{equation}\label{ghost_EOM2}
\nabla_\mu \left[(-4\pi\tilde{T}A_\partial^4 \beta_\partial+1) \nabla^\mu \phi\right] = m_\phi^2 \phi \ .
\end{equation}
for $A_\partial(\kappa)=e^{\beta_\partial \kappa/2}$. Alternatively, we can move the
negative sign to the left hand side of Eq.~\ref{ghost_EOM2} and obtain
a tachyonic linearized EOM
\begin{equation}\label{ghost_EOM3}
g^{tt} \partial_t^2\phi + \dots \approx  \left(-4\pi A_{\partial}^4 \beta_{\partial} \tilde{T} +1 \right)^{-1} \mphi^2 \phi 
\  .
\end{equation}
So, the two spontaneous scalarization theories behave similarly at the level of the
EOM.
Many of the conclusions for this action are similar to the theory of Eq.~\ref{st_action}.
When we need to distinguish the two forms of spontaneous scalarization, 
we will call the theory of Eq.~\ref{st_action} \emph{tachyon-based spontaneous scalarization}.

There are also important distinctions between tachyon- and ghost-based spontaneous scalarization.
The effective mass term $(-4\pi A_{\partial}^4 \beta_{\partial} \tilde{T} +1 )^{-1} \mphi^2$
can possibly diverge in ghost based spontaneous scalarization, leading to peculiar changes
in the NS structures. We will discuss this issue further in the discussion section. We 
consider ghost-based spontaneous scalarization as a purely classical modification 
to general relativity, which itself is a classical field theory. However, if one wishes to quantize
this theory, more care is needed compared to tachyon-based spontaneous
scalarization due to the particularly dangerous behavior of ghosts. Like the tachyon, it is also
possible, but not explicitly demonstrated, that the ghost is also regularized, i.e. the instability
shuts off as the ghost field grows~\cite{Ramazanoglu:2017yun}.
If this is the case, we might also have some improvement of the quantum
behavior as in ghost condensation~\cite{ArkaniHamed:2003uy}. A thorough study of this
question is beyond the scope of this paper.

We can also combine the two ideas to obtain a theory of \emph{ghost based
spontaneous vectorization} which has the action~\cite{Ramazanoglu:2017xbl}
\begin{align}\label{action_vt_ghost}
 \frac{1}{16\pi} &\int dV R -  \frac{1}{16\pi} \int dV \left[ F^{\mu\nu} F_{\mu\nu}  +2m_X^2 X^\mu X_\mu \right] \nonumber \\
 &+ S_m \left[\psi_m, A_F^2 g_{\mu \nu} \right], A_F =e^{\beta_F F^{\mu\nu} F_{\mu\nu}/4} \ ,
\end{align}
and the EOM
\begin{align} \label{eom_vt_ghost}
\nabla_\rho[(-4\pi A_F^4 \beta_F \tilde{T}\ +1) F^{\rho \mu}]& = m_X^2 X^\mu 
\end{align}

\emph{Spontaneously growing non-Abelian field theories}
have not appeared in the literature to the best of our knowledge. 
They can be formulated in a similar manner to spontaneous vectorization as
\begin{align}\label{action_ym0}
S=\phantom{+}&\frac{1}{16\pi} \int dV R
 \nonumber \\
-& \frac{1}{16\pi} \int dV (F^{a\mu\nu} F^a_{\mu\nu} + 2m_W W^a_\mu W^{a\mu}) \nonumber \\
+& S_m\left[\psi_m, A_{W}^2 g_{\mu \nu} \right] \ ,
\end{align}
where 
\begin{align}\label{YM0}
F^a_{\mu\nu} &= \nabla_\mu W^a_\nu - \nabla_\nu W^a_\mu + e f^{abc} W^b_\mu W^c_\nu \ .
\end{align}
$a,b,c$ label the vector potential $W^a_\mu$ and the generators $T^a$ of the
Lie algebra of the gauge group of the theory, $[T^a,T^b]=if^{abc}T^c$ is the defining
equation for the structure constants $f$, and $e$ is a coupling
constant~\cite{frampton2008gauge,Srednicki:2007qs}.
For example, if the gauge group is $SU(2)$ (a Yang-Mills theory),
$T^a$ are the Pauli matrices,
$f^{abc} =\epsilon^{abc}$ and $a=1,2,3$. 
\begin{align}\label{A_W}
A_W =\exp \left( \frac{\beta_W}{2} W^a_\mu W^{a\mu} \right)\ ,
\end{align}
in analogy to Eq.~\ref{action_vt}, or 
\begin{align}
A_W =\exp \left( \frac{\beta_{F,W}}{2} F^{a\mu\nu} F^a_{\mu\nu} \right)\ ,
\end{align}
in analogy to Eq.~\ref{action_vt_ghost} provide tachyon- or ghost-based versions of
the theory, respectively.

\section{Spontaneous Growth Through the Higgs Mechanism}\label{higgs_based}
Spontaneous vectorization in Eq.~\ref{action_vt} is very simply
analogous to spontaneous scalarization in Eq.~\ref{st_action} in terms of
mathematical form, and it is not surprising it was conceived as the
first generalization of spontaneous growth beyond scalars.\footnote{Eq.~\ref{action_vt}
was introduced earlier in a cosmological setting~\cite{BeltranJimenez:2013fca}}
However, this
action explicitly breaks the gauge symmetry
\begin{align}\label{gauge}
X_\mu \to X_\mu - \partial_\mu \lambda
\end{align}
that leaves the minimally coupled vector field action
\begin{align}\label{action_photon}
\frac{1}{16\pi} &\int dV R -  \frac{1}{16\pi} \int dV F^{\mu\nu} F_{\mu\nu}
\end{align}
invariant, where $\lambda$ is a scalar function. The term directly responsible for spontaneous growth, 
$A_X$, breaks the symmetry, hence it seems explicit symmetry breaking
is unavoidable in this formulation.\footnote{The mass terms $m_\phi$ and $m_X$ are
not necessary for tachyon-based spontaneous growth, they actually
inhibit it. However, $m_\phi$ is needed for agreement with observations~\cite{Ramazanoglu:2016kul},
and we expect a similar case for $m_X$~\cite{Ramazanoglu:2017xbl}.}
$A_F$ in ghost-based spontaneous
vectorization (Eq.~\ref{action_vt_ghost}) respects the symmetry in Eq.~\ref{gauge},
however the mass term $2m_X^2 X^\mu X_\mu$ is necessary in this
theory to have spontaneous growth in the EOM (Eq.~\ref{eom_vt_ghost}).
Hence, the gauge symmetry still has to be explicitly broken.

Why are we concerned with the breaking of this symmetry? 
First, vector fields we know to exist in nature
do have gauge symmetry in some form. Second, gauge symmetry is also important for
quantization in the non-Abelian case\footnote{Although note that an explicit mass term 
is still admissible as long as we remain below the strong coupling scale}
~\cite{frampton2008gauge,Srednicki:2007qs,Ruegg:2003ps}.
These concerns do not invalidate spontaneous vectorization
as formulated in Eq.~\ref{action_vt} as long as we view it as a classical
theory and do not identify $X^\mu$ with a known field in nature, i.e if it is
considered to be a yet-undiscovered field of gravitational interactions.
However, lack of symmetry makes this theory less ``natural'', and weakens
the motivation to study it.

Before we conceive a theory of spontaneous growth that respects the gauge 
symmetry, let us see how we can make a minimally coupled vector field under GR,
Eq.~\ref{action_photon}, massive using the famed Higgs
mechanism~\cite{frampton2008gauge,Srednicki:2007qs}. For
the simplest case, let us consider the action of a massless vector
together with a massless complex scalar
$\Phi$
\begin{align}\label{action_photon_higgs}
 \frac{1}{16\pi} &\int dV R -  \frac{1}{16\pi} \int dV F^{\mu\nu} F_{\mu\nu} \nonumber \\
+ \frac{1}{16\pi} &\int dV (2 \overline{D^\mu \Phi} D_\mu \Phi +2V(\Phib\Phi) )
\end{align}
Here
\begin{align}
D_\mu \Phi = (\nabla_\mu - i e X_\mu) \Phi
\end{align}
is the gauge invariant derivative, $e$ is a real coupling constant and
an over-bar denotes complex conjugation.
The action respects the gauge symmetry through the transformation
\begin{align}\label{gauge2}
X_\mu \to X_\mu - \nabla_\mu \lambda \ \ , \ \  \Phi \to e^{ie\lambda} \Phi \ .
\end{align}
Where is the mass term? At the most basic level, expanding $D_\mu$
brings a Lagrangian density
\begin{align}\label{action_photon_L}
\mathcal{L}_X =
-\frac{1}{4} F^{\mu\nu} F_{\mu\nu} - \frac{1}{2}(e^2\Phib\Phi) X^\mu X_\mu +\ \dots
\end{align}
Thus, $e^2\Phib\Phi$ behaves as an effective mass-square term. 
This does not mean much by itself since linear perturbations of $X_\mu$
are still massless around the solution with $X_\mu = 0 = \Phi$. However, the mechanism
is completed by the potential 
\begin{align}\label{higgs_V}
V(\Phib\Phi) = \frac{1}{2}\frac{m_0^2}{u^2}(u^2 - \Phib\Phi)^2
\end{align}
for real constants $u$ and $m_0$. The scalar behaves as a tachyon around $\Phi=0$
\begin{align}\label{eom_higgs00}
\Box \Phi = m_0^2(\Phib\Phi/u^2-1)\Phi +\ \dots
\end{align}
with linearized mass $im_0$, but it has a stable equilibrium for $|\Phi| = u$.
The stable configuration, whatever $\Phi$ we choose, provides a nonzero
value of $\Phib\Phi=u^2$. Even though the Lagrangian is symmetric, the ground state
of $\Phi$ is not, and this phenomena is called \emph{spontaneous symmetry breaking}.
Hence, for linear perturbations around the equilibrium values of the fields,
$X^\mu$ behaves as if it has mass $eu$. 

Since spontaneous scalarization in Eq.~\ref{st_action} and vectorization in
Eq.~\ref{action_vt} are based on mass generation in their
essence, Higgs mechanism immediately suggests a new type of spontaneous
growth. Remember that the conformal coupling had a mathematical form
similar to the mass potential term of the vector Lagrangian in Eq.~\ref{action_vt}.
Thus, we consider an action where the conformal scaling function has the form
that generates the $e^2u^2$ term in Eq.~\ref{action_photon_higgs} as
\begin{align}\label{action_vt_higgs0}
\phantom{+}&\frac{1}{16\pi} \int dV R - \frac{1}{16\pi} \int dV F^{\mu\nu} F_{\mu\nu} \nonumber \\
-&\frac{1}{16\pi}\int dV \big(2 \overline{D_\mu \Phi} D^\mu \Phi +2V(\Phib\Phi)\big) \nonumber \\
+& S_m \left[\psi_m, A_H^2 g_{\mu \nu} \right] \ ,
\end{align}
where 
\begin{align}\label{A_H0}
A_H = \exp \left( \frac{\beta_H}{2} \overline{D_\mu \Phi} D^\mu \Phi \right) \ .
\end{align}
Variation of the action leads to
\begin{align}\label{eom_higgs0}
G_{\mu\nu} &= 8\pi T_{\mu\nu}+ T^{\Phi,X}_{\mu\nu} -g_{\mu\nu} V(\Phib\Phi) \nonumber \\
\nabla^\nu F_{\nu\mu} &=\gamma_H^{-1}
(e^2 \Phib\Phi\ X_\mu + J^\Phi_\mu)  \\
\mathcal{D}^2 \Phi
&=\big[ \gamma_H^{-1}m_0^2(\frac{\Phib\Phi}{u^2}-1)+e^2X_\mu X^\mu +ieX^\mu \nabla_\mu]\Phi  \nonumber 
\end{align}
where
\begin{align}\label{higgs_source}
\mathcal{D}^2 \Phi &= \gamma_H^{-1}\nabla_\mu[\gamma_H (\nabla^\mu -ieX^\mu) \Phi] \nonumber \\
\gamma_H &= -4\pi \beta_H \tilde{T} A_H^4 +1 \nonumber \\
T_{\mu\nu} & = A_H^2 \tilde{T}_{\mu \nu}
- \beta_H \tilde{T} A_H^4 \overline{D_{(\mu} \Phi} D_{\nu)} \Phi  \nonumber \\
T^{\Phi,X}_{\mu\nu} &= 2 \nabla_{(\mu}\Phib \nabla_{\nu)} \Phi 
-g_{\mu\nu}  g^{\rho \sigma} \nabla_\rho \Phib \nabla_\sigma \Phi  \nonumber \\
&\ \ +2F_{\mu\rho} F_\nu{}^{\rho} -\frac{1}{2} g_{\mu\nu} F_{\rho \sigma} F^{\rho\sigma}  \\
&\ \ +2e^2 \Phib\Phi \left( X_\mu X_\nu - \frac{1}{2} g_{\mu\nu}  
g^{\rho \sigma} X_\rho X_\sigma \right) \nonumber \\
&\ \ +2J_{\mu}^\Phi X_{\nu}+2J_{\nu}^\Phi X_{\mu} - 2g_{\mu\nu}  g^{\rho \sigma} J^\Phi_\rho X_\sigma \nonumber \\
J^\Phi_{\mu} &= \frac{ie}{2} (\Phib\nabla_\mu \Phi - \Phi \nabla_\mu \Phib ) \nonumber
\end{align}
If matter is not very relativistic, $\tilde{T}= -\rt+3\pt\approx -\rt<0$,
and appropriate negative values of $\beta_H$ provides a negative 
effective vector mass-square $m_{\rm eff}^2 = (-4\pi \beta_H \tilde{T} A_H^4 +1)\ 
e^2 \Phib\Phi$. This means the linearized EOM is tachyonic for 
$4\pi \beta_H \tilde{T} A_H^4>1$
thanks to the non-zero value of $\Phib\Phi$.
The construction respects the gauge symmetry (Eq.~\ref{gauge2}) in its explicit form.
We call this theory \emph{Higgs-based spontaneous vectorization}.

There is one issue that needs elaboration in the above explanation.
Remember that the pure Higgs EOM,
Eq.~\ref{eom_higgs00}, is already tachyonic around $\Phi=0$. However, the nonminimal 
coupling terms
bring another factor in front of $m_0^2$ in
Eq.~\ref{eom_higgs0}, $(-4\pi \beta_H \tilde{T} A_H^4 +1)^{-1}$, which is negative if
we want $X^\mu$ to spontaneously grow. Thus, the tachyonic instability of $\Phi$
is eliminated wherever $X^\mu$ is tachyonic. This might suggest at a 
first look that $\Phib\Phi$ does not attain nonzero equilibrium values since
the scalar never spontaneously grows from $\Phi=0$, hence the effective
tachyonic vector mass term $e^2\Phib\Phi$ also vanishes, meaning there is no
Higgs mechanism for $\Phi$ or spontaneous growth of $X^\mu$.
This is \emph{not} the case. $\Phi$ is continuous, and if it attains a
nonzero value due to an instability in any region of spacetime, it
generically attains nonzero values everywhere. For example, in the
spontaneous scalarization theory of Eq.~\ref{st_action} the tachyon 
strictly lives inside the NS, but the scalar field is nonzero everywhere, not just
inside the NS. $\phi$ merely dies off away from the star, and vanishes only
at spatial infinity for generic solutions. This means, since the Higgs
mechanism is intact in vacuum in Eq.~\ref{eom_higgs0}, $\Phib\Phi$
still attains nonzero values where there is matter even though $\Phi$ is not tachyonic
there. Overall, $X^\mu$ retains its tachyonic modes and spontaneously grows.

The $X^\mu$ terms in the EOM (Eq.~\ref{eom_higgs0}) also behave as
mass-square terms at the non-perturbative level since $X^\mu$ also attains
non-zero equilibrium values when it grows spontaneously. These terms might enhance
or inhibit the instability of $\Phi$ around its zero value depending on their overall sign, but
the same continuity argument can be applied to them as well.
At equilibrium, generically $\Phib\Phi \neq 0$ which is enough for the spontaneous growth of $X^\mu$.

A second form of gauge symmetry-respecting spontaneous growth theory exists
which utilizes spontaneous growth itself to provide a non-zero equilibrium value for the scalar. 
The role of $V(\Phib\Phi)$ is shifting the equilibrium value of $\Phi$ away from zero,
but tachyon-based spontaneous scalarization already achieves 
this locally inside NSs which is where we need it. 
Consider the action 
\begin{align}\label{action_vt_higgs}
\phantom{+}&\frac{1}{16\pi} \int dV R - \frac{1}{16\pi} \int dV F^{\mu\nu} F_{\mu\nu} \nonumber \\
-&\frac{1}{16\pi}\int dV\ 2 \overline{D_\mu \Phi} D^\mu \Phi \nonumber \\
+& S_m \left[\psi_m, A_H^2 g_{\mu \nu} \right] 
\end{align}
where 
\begin{align}\label{A_H}
A_H = \exp \left( \frac{\beta_H}{2} \overline{D_\mu \Phi} D^\mu \Phi\ -\frac{\beta_H m_\Phi^2}{2} \Phib\Phi \right) \ ,
\end{align}
and $\beta_H$, $m_\Phi$ are real constant parameters of the theory.
The EOMs are
\begin{align}\label{eom_higgs}
G_{\mu\nu} &= 8\pi T_{\mu\nu}+ T^{\Phi,X}_{\mu\nu} \nonumber\\
\nabla^\nu F_{\nu\mu} &=(-4\pi \beta_H \tilde{T} A_H^4 +1) 
(e^2 \Phib\Phi\ X_\mu + J^\Phi_\mu)  \\
\mathcal{D}^2 \Phi
&=\big[ \gamma_H^{-1}(1-\gamma_H) m_\Phi^2+e^2X_\mu X^\mu +ieX^\mu \nabla_\mu]\Phi  \nonumber 
\end{align}
where $T_{\mu\nu}$, $T^{\Phi,X}_{\mu\nu}$, $J^\Phi$, $\mathcal{D}^2\Phi$ and $\gamma_H$
are as in Eq.~\ref{higgs_source}
but with the new definition of $A_H$ in Eq.~\ref{A_H}.

We can repeat our previous line of thought and easily see that Eq.~\ref{eom_higgs}
shows spontaneous growth of $\Phi$ and $X^\mu$.
Moreover, this time the mass-square term on the first line of the 
EOM for $\Phi$ is manifestly tachyonic for $4\pi \beta_H \tilde{T} A_H^4>1$.
For example, when $4\pi \beta_H m_\Phi^2 \tilde{T} A_H^4 \ll -1$, this
terms becomes $-m_\Phi^2$ in the leading order, hence we can see that
the nonzero equilibrium value of $\Phib\Phi$ is attained by its own spontaneous
growth in the presence of matter, but not as a side effect of tachyonic modes elsewhere
in the spacetime.

An interesting aspect of Eq.~\ref{A_H} is that the part that normally causes tachyon-based
spontaneous growth, $-\beta_H m_\Phi^2 \Phib\Phi/2$ has the wrong sign,
it generates the leftmost $1-\gamma_H$ term in the EOM of $\Phi$,
and does \emph{not} cause spontaneous growth by itself.
However, the $\beta_H \overline{D_\mu \Phi} D_\mu \Phi /2$
term in $A_H$ causes ghost based spontaneous scalarization through the factor
$\gamma_H^{-1}$ at the same time, and
only the combination of the two provides the negative mass-square term
of $\Phi$ in Eq.~\ref{eom_higgs}.

To summarize, $\Phi$ spontaneously grows in Eq.~\ref{action_vt_higgs}
if there is matter, which in turn leads to the spontaneous
growth of $X^\mu$ through the tachyonic term $m_{\rm eff}^2 = (-4\pi \beta_H \tilde{T} A_H^4 +1)\ 
e^2 \Phib\Phi$. The construction again respects the gauge symmetry, which is only
spontaneously broken through the non-zero equilibrium value of $\Phi$.
One difference from the theory in Eq.~\ref{action_vt_higgs0} is that $X^\mu$
becomes asymptotically massless away from the matter in this case, while it
is massive in the latter.

Eq.~\ref{action_vt_higgs} is reminiscent of the recently introduced idea of
\emph{gravitational Higgs mechanism} where spontaneous scalarization leads to nonzero
equilibrium value for the scalar which in turn generates real mass values for gauge
fields~\cite{Coates:2016ktu}. Due to this similarity,
we call this phenomena \emph{gravitational Higgs-based
spontaneous vectorization}.
However,  we use spontaneous scalarization terms to incite the
growth of the scalar field itself only, and the tachyonic modes of the gauge field
$X^\mu$ are due to the derivative terms in Eq.~\ref{A_H} in contrast to~\cite{Coates:2016ktu}.

We should also emphasize that $\Phi$ in Eq.~\ref{action_vt_higgs0}
is not the Higgs field of the Standard Model. $\Phi$
and $X^\mu$ can both be better viewed as fundamental degrees of freedom 
associated with gravity that are hitherto unobserved. 
However, their possible association to the Standard Model is an interesting
research question to study.\footnote{We are overlooking
the fact that the Higgs field of the Standard Model is not a single complex scalar, which
is not central to our point.}

Once the fundamental elements of gauge symmetry-respecting spontaneous growth are
in place, we can also apply it to more complex field theories. Without going into details,
a theory of spontaneously growing non-Abelian gauge fields $W^a_\mu$, e.g. Yang-Mills
theory, is given by the action
\begin{align}\label{action_ym_higgs}
\phantom{+}&\frac{1}{16\pi} \int dV R
-\frac{1}{16\pi}\int dV  \big( 2(D_{\mu} \Phi)^\dagger D^\mu \Phi +2V(\Phi^\dagger\Phi) \big)
 \nonumber \\
-& \frac{1}{16\pi} \int dV F^{a\mu\nu} F^a_{\mu\nu}
+\ S_m\left[\psi_m, A_{YM}^2 g_{\mu \nu} \right] \ ,
\end{align}
where all quantities are defined as in Eq.~\ref{action_ym0} and $^\dagger$
indicated a Hermitian conjugate. The Higgs field is now a multidimensional object 
that can be acted upon by $T^a$, and
\begin{align}\label{YM}
D_\mu\Phi &=\nabla_\mu \Phi - ie W_\mu^b T^b \Phi \ .
\end{align} 
In analogy to Eq.~\ref{action_vt_higgs0}
\begin{align}
V(\Phi^\dagger \Phi) = \frac{1}{2}\frac{m_0^2}{u^2}(u^2 - \Phi^\dagger\Phi)^2 \ 
\end{align}
and
\begin{align}\label{A_YM}
A_{YM} = \exp \left( \frac{\beta_{YM}}{2} (D_{\mu} \Phi)^\dagger D^\mu \Phi  \right) \ .
\end{align}

Similarly, we can have the alternative formulation of gravitational Higgs-based
spontaneous growth of $W^a_\mu$ if $V(\Phi^\dagger\Phi)=0$, and
\begin{align}\label{A_YM0}
A_{YM} = \exp \left( \frac{\beta_{YM}}{2} (D_{\mu} \Phi)^\dagger D^\mu \Phi 
-\frac{\beta_{YM} m_\Phi^2}{2} \Phi^\dagger \Phi \right) 
\end{align}
for appropriate constants $\beta_H$ and $m_\Phi$. 

We will not attempt to numerically find the structures of NSs with spontaneously grown
Abelian or non-Abelian gauge fields in this study, but will discuss certain
aspects of such solutions in the final section. EOMs can be easily reduced to a
set of coupled ordinary differential equations similar to Tolman-Oppenheimer-Volkoff
equations following standard algebra~\cite{Ramazanoglu:2017yun,Franchini:2017zzx}.


\section{Discussion}\label{discussions}
We presented a new form of spontaneous growth mechanism in gravity
for gauge fields where the Higgs mechanism is employed
to avoid explicit symmetry breaking in the action.  
Using the Higgs mechanism is a natural choice since the prototypical
spontaneous growth theory, spontaneous scalarization, also relies
on a mass generation mechanism much like Higgs. We utilize this
fact to devise two different theories where the nonzero equilibrium
value of the Higgs scalar can arise from a potential term or spontaneous
scalarization of the Higgs field.

The Higgs mechanism also leads to modifications 
in the EOM of the Higgs scalar itself.
The effective mass of $\Phi$ necessarily diverges in Eq.~\ref{eom_higgs0}
and~\ref{eom_higgs} for astrophysical systems due to the
$(1-4\pi \beta_H \tilde{T} A_H^4)^{-1}$ terms. This is because $\tilde{T}$ vanishes
in vacuum and consequently $1-4\pi \beta_H \tilde{T} A_H^4=1$ outside a
star, whereas $(1-4\pi \beta_H \tilde{T} A_H^4)<0$ has to be satisfied inside if
there is to be a spontaneous growth of $X^\mu$ at all. This means
$(1-4\pi \beta_H \tilde{T} A_H^4)^{-1}$ necessarily diverges at some points
inside a star that experiences Higgs-based spontaneous vectorization.

Such divergent terms are familiar from ghost-based spontaneous growth
theories where they have been examined in more detail~\cite{Ramazanoglu:2017yun}.
Even though there are infinities in the EOM,
physical quantities are still finite and continuous, but there are
cusps in the density profiles of stars if they spontaneously grow fields 
this way.
It is likely that such features are relatively easy targets for observation, and 
will lead to strong restrictions
in the parameter space of Higgs-based spontaneous growth theories. 
Strictly tachyon based theories in Eq.~\ref{action_vt}
and Eq.~\ref{action_ym0}-\ref{A_W} do not have this problem, and likely lead to
more regular NS structures. Our current level of knowledge suggests that gauge symmetry and smooth
NS structures cannot be achieved in the same theory.
Future studies of NSs and their mergers will be the final arbiter of the merit
of these two types of theories relative to each other and in comparison to GR. 

One difference between the two Higgs mechanisms we investigated is that plain
Higgs (Eq.~\ref{action_vt_higgs0}) provides a nonzero $X^\mu$ mass everywhere,
whereas gravitational Higgs  (Eq.~\ref{action_vt_higgs}) does not
generate mass as one goes away from the spontaneously growing field, i.e. the NS.
The parameter space for spontaneous scalarization is known to be severely 
restricted for massless scalars due to measurements from non-merging binary
systems~\cite{Ramazanoglu:2016kul,2013Sci...340..448A}.
This in turn strongly suggests that spontaneous growth
of massless Abelian or non-Abelian gauge theories are also at odds with 
existing tests of gravity.
Hence, we can say that the gravitational Higgs-based spontaneous growth is
unfavored, but not ruled out, by observations. We leave the quantitative analysis
of this issue to future studies.

We have not discussed the regularization of the instabilities we have shown to exist
in gauge field theories. Namely, an instability is not a desired feature in a theory of
nature unless it stops growing by additional mechanisms. The fourth order
term of the Higgs potential $V$ that makes $\Phi$ reach equilibrium at finite values,
and the suppressive effect of the $A^{\beta\phi^2/2}$ on the tachyonic terms of 
Eq.~\ref{scalar_eom} as $\phi$ grows are such mechanisms.
The arguments in the exponent of
$A_H$ are not negative definite in either of the two Higgs-based spontaneous
growths. Thus, it is not clear
which part of the more complicated EOMs Eq.~\ref{eom_higgs0} and~\ref{eom_higgs} 
can make sure that the instabilities eventually shut off and
lead to stable gauge field clouds.
Understanding the stability of the final NS configurations with 
these growth mechanisms requires detailed numerical studies. It is noteworthy 
that such questions about the fate of the instability are not easily answered for any
generalization of tachyon-based spontaneous scalarization. This statement 
is true for tachyon-based spontaneous vectorization and ghost-based spontaneous
scalarization theories as well for which numerical solutions 
have been constructed~\cite{Ramazanoglu:2017xbl,Ramazanoglu:2017yun}.

One may argue that mathematically $\overline{D_\mu \Phi} D^\mu \Phi$ simply
hides an $X_\mu X^\mu$ term, and is not really different from tachyon based spontaneous
vectorization. We can partially embrace such a comment since this is the core of
the Higgs mechanism: it contains a hidden effective mass term, but in such a way that the gauge 
symmetry is intact in the action. Moreover, the Higgs mechanism also provides
a unified mechanism that generates mass for various fields at the same time
as in the Standard Model. This means one can actually devise theories
where all fields are spontaneously grown in the presence of matter by a single
Higgs kinetic term in the nonminimal matter coupling. However, we remind that
we do not identify any of the spontaneously growing fields in this study with known
particles of the Standard Model.

Higgs fields are also the source of the mass of spinor fields in the Standard Model,
but we have not discussed a Higgs-based spontaneous spinorization theory.
This is because spinor masses are generated through the Yukawa mechanism
through terms that schematically look like $\psib \Phi \psi$ that are linear, not quadratic, in
$\Phi$~\cite{frampton2008gauge,Srednicki:2007qs}.
Changing the sign of this term does not provide an instability unlike the gauge
fields we considered which have $\Phib\Phi$ terms. Spontaneous spinorization theories 
has been introduced by other means~\cite{Ramazanoglu:2018hwk}, but we believe
it would be valuable to find a Higgs-based formulation for spinors as well to
demonstrate the universality of the Higgs field to generate tachyonic instabilities
as well as mass.

Symmetry is a central concept in modern physics. 
The main aim of this study was a new formulation of spontaneous growth which 
would better reflect the fundamental symmetries of nature. We believe the
theories we introduced constitute important near-future targets for physics of strong gravity 
and gravitational wave science.
 
\acknowledgments
We are grateful to Bayram Tekin for stimulating discussions.
The author is supported by
Grant No. 117F295 of the Scientific and Technological
Research Council of Turkey (T\"{U}B\.{I}TAK).We would like to
acknowledge networking and travel support by the COST
Action CA16104.

\bibliography{/Users/fethimubin/research/papers/references_all}

\end{document}